\documentclass{article}
\usepackage{bigsky2007}
\usepackage{graphicx}
\frompage{000} \topage{000}                                              

\usepackage{graphicx}
\def\rightmark{THERMINATOR simulations and PHENIX images}
\def\leftmark{R. V\'ertesi for the PHENIX Collaboration}

\newcommand{\bl}[1]{\begin{equation}\label{#1}} 
\newcommand{\ee}{\end{equation}}                
\newcommand{\rec}[1]{\frac{1}{#1}}              
\newcommand{\z}[1]{\left({#1}\right)}           
\renewcommand{\v}[1]{\mathbf{#1}}               
\newcommand{\ud}{\,\mathrm{d}}                  
\def\r12{\ensuremath{r_{12}}}
\def\kt{\ensuremath{k_{\mathrm{t}}}}

\def\Sr{\ensuremath{S(\r12)\;}}

\newcommand{\pT}{\ensuremath{p_T}}
\newcommand{\kT}{\ensuremath{k_T}}
\newcommand{\rhomax}{\ensuremath{\rho_{max}}}
\newcommand{\MeV}{\ensuremath{\mathrm{MeV}}}

\newcommand{\pip}{\ensuremath{\pi^\mathrm{+}}}
\newcommand{\pim}{\ensuremath{\pi^\mathrm{-}}}
\newcommand{\kTlo}{\ensuremath{200\ \MeV<\kT<360\ \MeV}}
\newcommand{\kThi}{\ensuremath{480\ \MeV<\kT<600\ \MeV}}

\title{THERMINATOR simulations and PHENIX images of a heavy tail of particle emission in 200 GeV Au+Au collisions
}
\authors{
{R. V\'ertesi$^{1,2}$ for the PHENIX Collaboration
}\\[2.812mm]
{\normalsize
\hspace*{-8pt}$^1$ KFKI Research Institute for Particle and Nuclear Physics, \\
Budapest, Hungary\\[0.2ex]
\hspace*{-8pt}$^2$ University of Debrecen, \\
Debrecen, Hungary\\[0.2ex] %
}}

\abstract{Correlation between emitted particles from a thermalized medium carry important
information about the space-time extent and the dynamics of the particle emitting source. 
Pion emission source functions $S{(\bf{r})}$
have been measured in PHENIX using an imaging technique, and it has been found that it
contains a heavy (power-law) tail.
It leads us to the assumption that there is a halo of secondary pions, surrounding
the core system that undergoes the hydrodynamical evolution. 
THERMINATOR, a Monte Carlo event generator
designed for studies in relativistic heavy-ion collisions is used to
model and investigate the influence of resonance decays on the tail of particle emission source. 
It lacks the implementation of rescattering, which, as we
have learnt from previous Hadron Rescattering Code simulations, is a possible explanation of
the power-law tail. Our studies also have shown that none of the pion
sources alone are responsible for the tail: either the resonance decays or rescattering
or a combination of these effects is a possible cause of the heavy tail.}


\begin{document}

\maketitle
\setcounter{page}{1}

\section{Introduction}\label{lbl:intro}

Correlation between identical pions emitted from a thermalized source with 
similar momenta is due to the bosonic nature of the pions. The phenomenon is 
often referred to as Bose-Einstein Correlation (BEC). The correlation functions carry
important information about the space-time extent and the dynamics of the particle
emitting source. R.~Hanbury Brown and R.~Q.~Twiss used the bosonic correlation of particles 
to measure the angular diameter of distant stars~\cite{HBT}. BEC, or the HBT effect has
become an important tool in high energy physics. In most heavy ion experiments there
is enough statistics to measure the two- and three-particle correlation function 
of charged pions. We define the correlation function $C(k_1,k_2)$ as
\bl{eq:Cdef} C(k_1,k_2)=\frac{N_2(k_1,k_2)}{N_1(k_1)N_1(k_2)} .
\ee Here $k_1$ and $k_2$ are the four-momenta of the emitted
particles, $N_1(k)$ is the single particle momentum distribution,
$N_2(k_1,k_2)$ is the two-particle momentum distribution. To raise
the statistics, one integrates over some components of $k_1$,
$k_2$. From the experimental point of view the adopted definition
is \bl{eq:Cqdef} C(q)=\frac{N_{corr}(q)}{N_{mix}(q)} , \ee the
number of pairs with $q$ relative momentum from the same event
divided by the number from mixed events. This is measured directly
in experiments, and it must not to be confused with the ideal
Bose-Einstein correlation function without any final state
effects, $C_0(q)$. In our notation $q=\sqrt{-(k_1-k_2)^2}$ is the
invariant relative momentum of the pair, which is equal to
$2k_{12}$ in the pairs c.m.s. frame. The definition of $k_{12}$
is: $k_{12}=\rec{2}\left|\v{k}_1-\v{k}_2\right|$.

In the approximate case when multi-particle correlation effects
are negligible, the BEC has good theoretical description even in
the case when final state interactions (FSI) play important role.
The two-particle correlation function of the pions can be calculated from the
source function as follows \cite{Koonin}: \bl{eq:PPGinteq}
C(q)-1=\rec{2}\int\ud^3\v{r}\z{\left|\Phi_C\z{q,\v{r}}\right|^2-1}S\z{\v{r}}
, \ee where $\Phi_C(q,\v{r})$ is the symmetrized final state
two-particle outgoing wave function. In case of
Coulomb-interaction, this has an analytic form~\cite{Messiah,Alt}. With
the imaging method (by inverting this integral equation) one can
determine the $S(\v{r})$ source function. If we assume spherical
symmetry, then $S(\v{r})$ depends only on $r:=|\v{r}|$.

\section{PHENIX Source Imaging}\label{lbl:img}
\begin{figure}[htb]
  \begin{tabular}{cc}
    \begin{minipage}{50mm}
      \includegraphics[width=60mm]{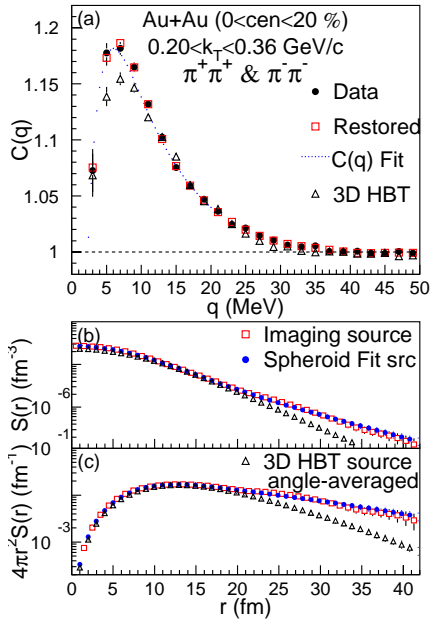}
    \end{minipage}
    &
    \begin{minipage}{75mm}
      \caption[]{\it (a) Correlation function, C(q), (full circles) for
      \pip\pip{} and \pim\pim pairs (\kTlo{}) produced in the central Au+Au collisions of
      0\%-20\% centrality. Open squares represent the restored
      correlation from the imaging technique. The dotted line represents
      the result of direct correlation fitting. The open triangles
      represent the 1D angle-averaged correlation of the 3D correlation
      function. (b) and (c) show the 1D source function $S(r)$ and
      $4\pi{}r^2S(r)$, respectively, from imaging (open squares),
      Spheroid fit to the correlation function (full circles), and
      angle-averaging of the 3D-Gaussian source function (open
      triangles). Systematic errors are less than the size of the data
      points.} \label{fig:ppg052_1}
    \end{minipage}
  \end{tabular}
\end{figure}

Emission source functions have been extracted from correlation
functions constructed from charged pions produced at mid-rapidity
in Au+Au collisions at $\sqrt{s_\mathrm{NN}}=200\
\mathrm{GeV}$ \cite{PPG052}. A model-independent imaging technique
developed by Brown and Danielewicz \cite{ImgSrc} has been carried out (See
Fig.~\ref{fig:ppg052_1}). The source parameters extracted from
low \kT~PHENIX data give the indication of a long tail, which
differs from that of a Gaussian distribution.

The aim of this talk was to present our investigations carried out
with THERMINATOR and HRC Monte Carlo simulatons, that concluded in
possible kinematic explanations of the long-range components of
the radial source function seen in PHENIX.

\section{Hadron Rescattering}\label{lbl:hrc}
It has been shown \cite{Humanic:2002iw,Humanic:2006ve} that
calculations based on a hadronic rescattering model agree
reasonably well with experimental results on single particle
spectra, elliptic flow, HBT radii in Au+Au collisions at RHIC
energies. The model we use here to compare PHENIX data with is Tom
Humanic's Hadronic Rescattering Calculations. The model deals with the 8
most abundant resonances and simulates their decays and rescattering.

Simulated source of charged pions is discussed in three groups, 
determined by time of creation. The {\it core} consists 
of the primordial particles
and the decay products of the short lived resonances. The secondary
pion sources of long lived resonances are referred to as the {\it
halo}, while the decay products of the $\omega(782)$ meson make up
a separate class (see Table \ref{tab:pisrc}).

\begin{table}[h!] 
\vspace*{-12pt}
\caption[]{Components of charged pion sources in the HRC simulation as grouped in this discussion.}
\label{tab:pisrc}
\vspace*{-14pt}
\begin{center}
\begin{tabular}{|l|l|l|}
\hline
Group & Lifetime & Components in HRC \\
\hline
\hline
core & $<20~\mathrm{fm/c}$ & primordial $\pi^\pm$, $\rho, \Delta, \mathrm{K}^\star$ \\
\hline
omega & $\approx 23~\mathrm{fm/c}$ & $\omega(782)$ \\
\hline
halo & $>25~\mathrm{fm/c}$ & $\Lambda, \Phi, \eta, \eta\prime$ \\
\hline 
\end{tabular}
\end{center}
\end{table}

\begin{figure}[htb]
\begin{center}
  \includegraphics[width=55mm]{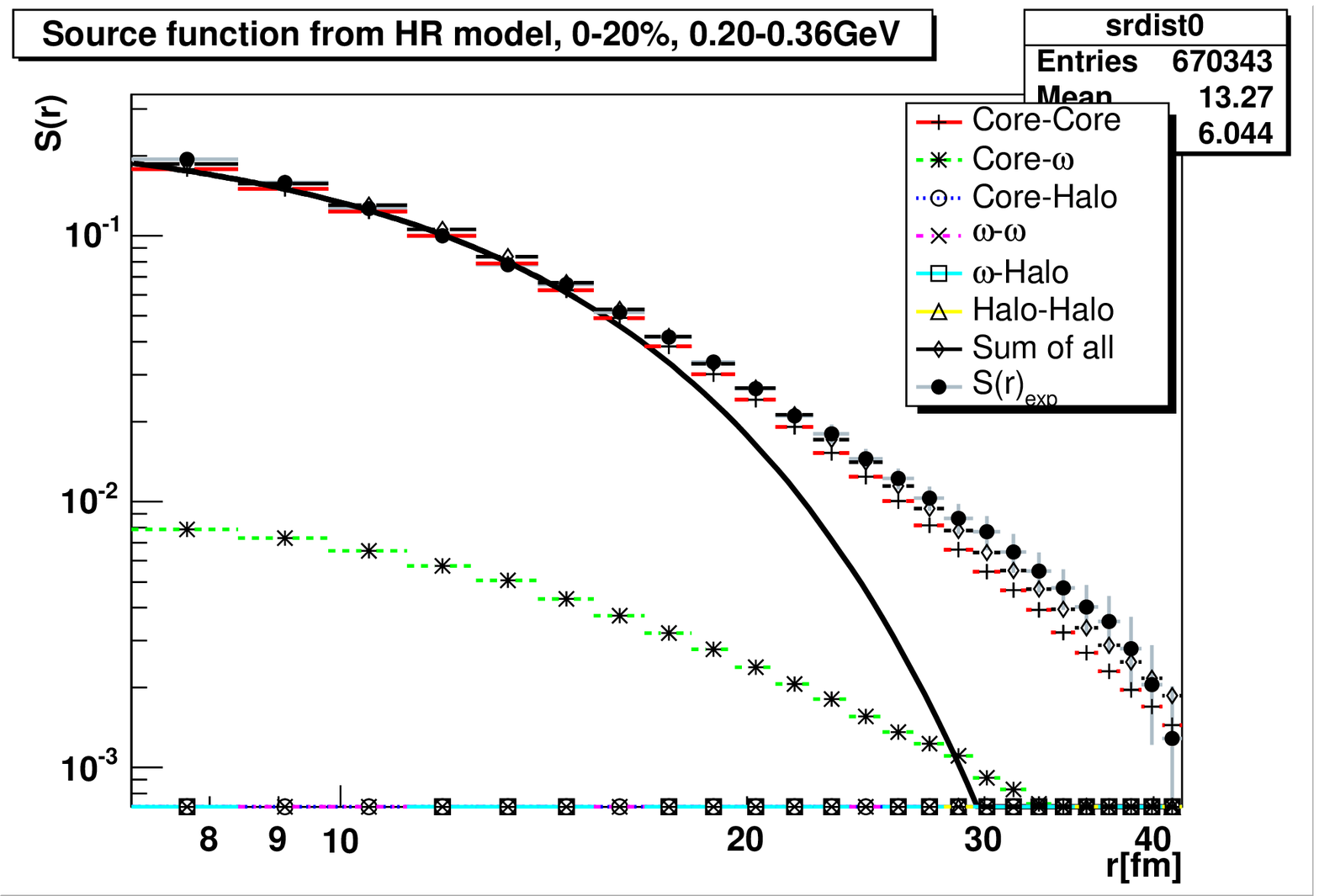}
  \includegraphics[width=55mm]{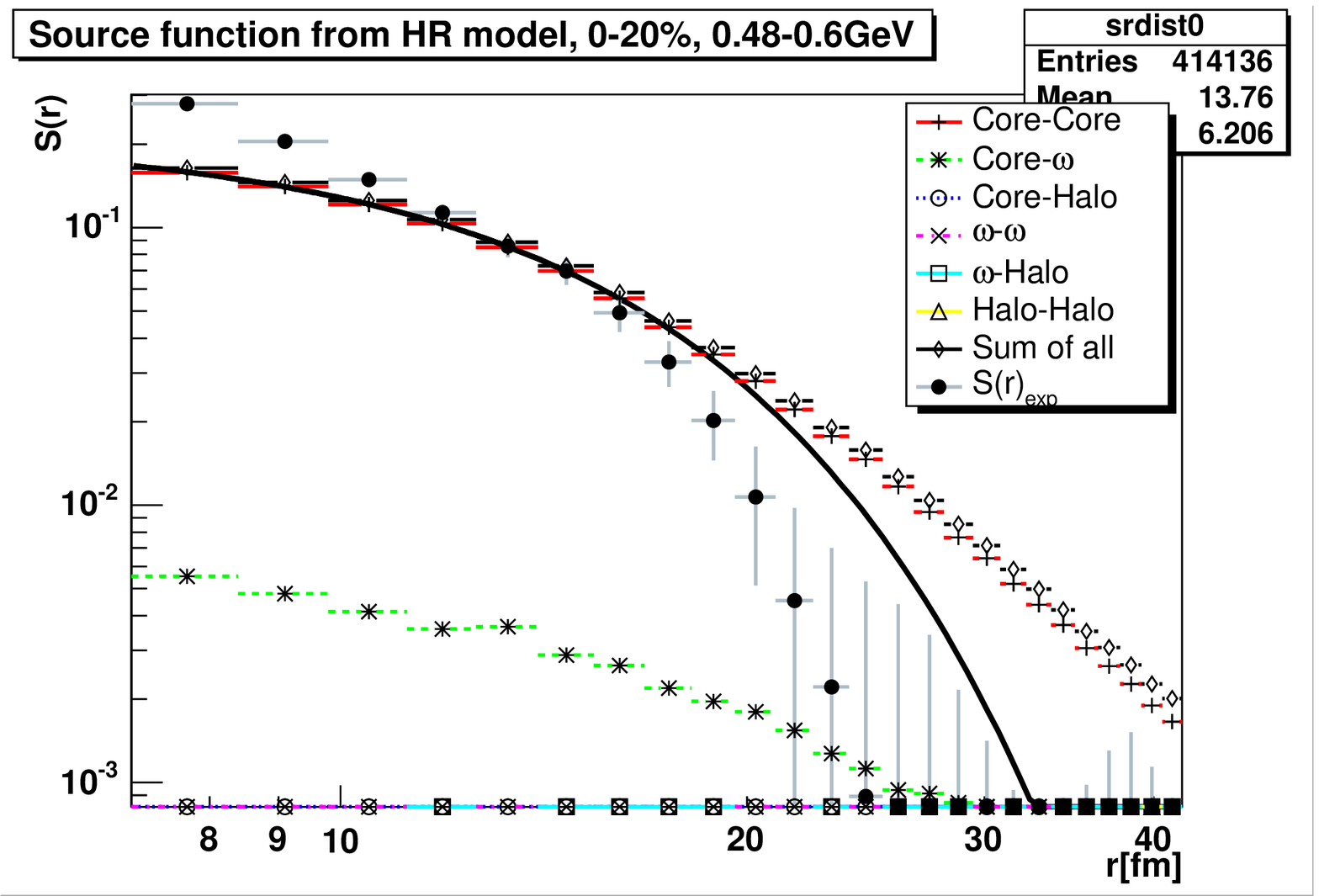}
  \\
  \includegraphics[width=55mm]{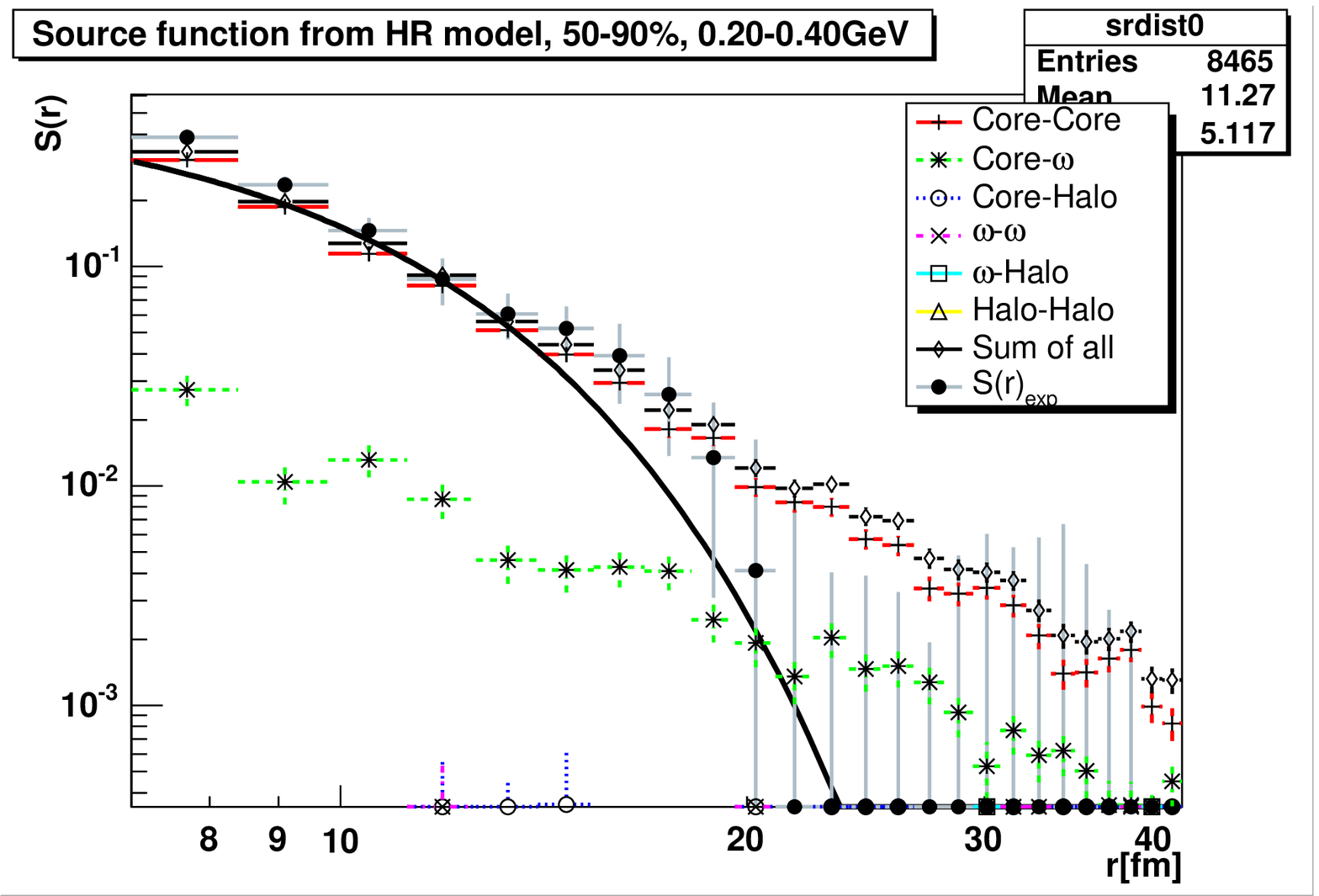}
  \caption[]{\it Pion pair source measured in PHENIX and HRC simulation for central
    ($0-20\%$) \kTlo~collisions (top left); high \kT~(\kThi) from the 20\% most
    central Au+Au midy collisions (top right); low kT (\kTlo) from
    peripheral collisions ($40-90\%$) (bottom center).
    Solid dots represent data on each figure. 
    Black crosses are simulation, while color online symbols stand for different
    components of the simulated source as explained on the legend. 
    The Gauss fit to $S(r<15~\mathrm{fm})$ is shown as a continuous line.}
\label{fig:hrc}
\end{center}
\end{figure}

On Fig.\ \ref{fig:hrc}, note the excellent agreement between
PHENIX data and HRC simulation in the central, low $\mathrm{k_T}$
case. The agreement is also acceptable in the full range for the
peripheral low $\mathrm{k_T}$ case. However, HRC fails to describe
the $\mathrm{k_T}$ dependence of the source.

Further investigations have shown that the $\omega(782)$ does not
account for the heavy tail in the simulation. Core-core pairs
themselves have a power-law tail in the \Sr distribution. In the
HRC code, rescattering goes on until it self-quenches. The mean
free path increases strongly as the system expands, and
rescattering in such a time dependent mean free path system
corresponds to an anomalous diffusion. Anomalous diffusion is known
to lead to power-law tailed distribution, as stated by a generalization 
of the Central Limit Theorem \cite{Metzler_Klafter,0702032}.

\section{Single Freezeout with THERMINATOR}

THERMINATOR---A Thermal Heavy Ion Generator \cite{Therm}, based on
the Cracow Single Freezeout Model \cite{Cracow} was used for the
following simulations. In this model, freezeout of the partonic matter occurs
on a single space-time surface, governed by universal thermodynamic
parameters. Particle distribution is determined by the
thermodynamical equilibrium. Particles propagate freely and the
decay of 385 resonances is fully implemented. In opposition to the HRC simulation, 
rescattering effects are neglected here.

\subsection{PHENIX source simulated}

THERMINATOR simulation provides a flat rapidity distribution.
Several checks were done to verify that a realistic distribution
and $p_\mathrm{T}$ spectrum of particles is generated.

Centrality can not be directly set in THERMINATOR, but through the
parameters $\rhomax$ and $\tau$, where $\rhomax$ stands for
a radius of a boost invariant axially symmetric cylinder, while 
$\tau$ denotes the proper time of a simultaneous freezeout and hadronization
\cite{Cent}. Although fitted parameters for the spectra of all four RHIC 
experiments including PHENIX were published, PHENIX \pT\ spectra is not well 
described by THERMINATOR, being primarily developed for, and tuned to STAR.
Thus we used the parameters corresponding to STAR spectrum fits.
This indicates that THERMINATOR might be able to show much better
agreement with data with further tuning. The central data region
$0-20\%$ has been reproduced using simulated events from the three
most central regimes in proportion to their span. The $40-90\%$
regime was substituted with equal number of events simulated in the regimes
of $50-60\%$ and $70-80\%$.

\begin{figure}[htb]
\begin{center}
  \includegraphics[width=55mm]{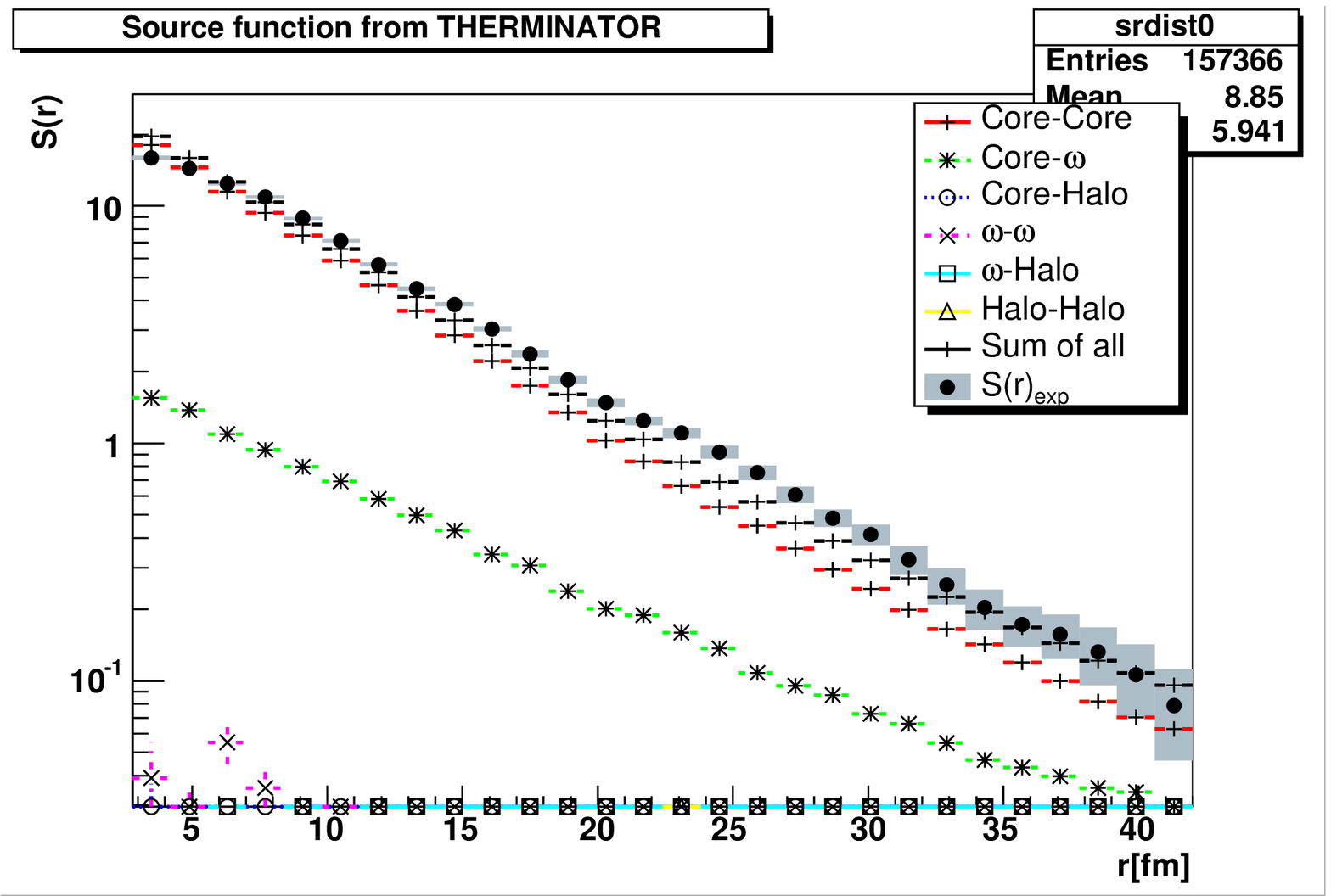}
  \includegraphics[width=55mm]{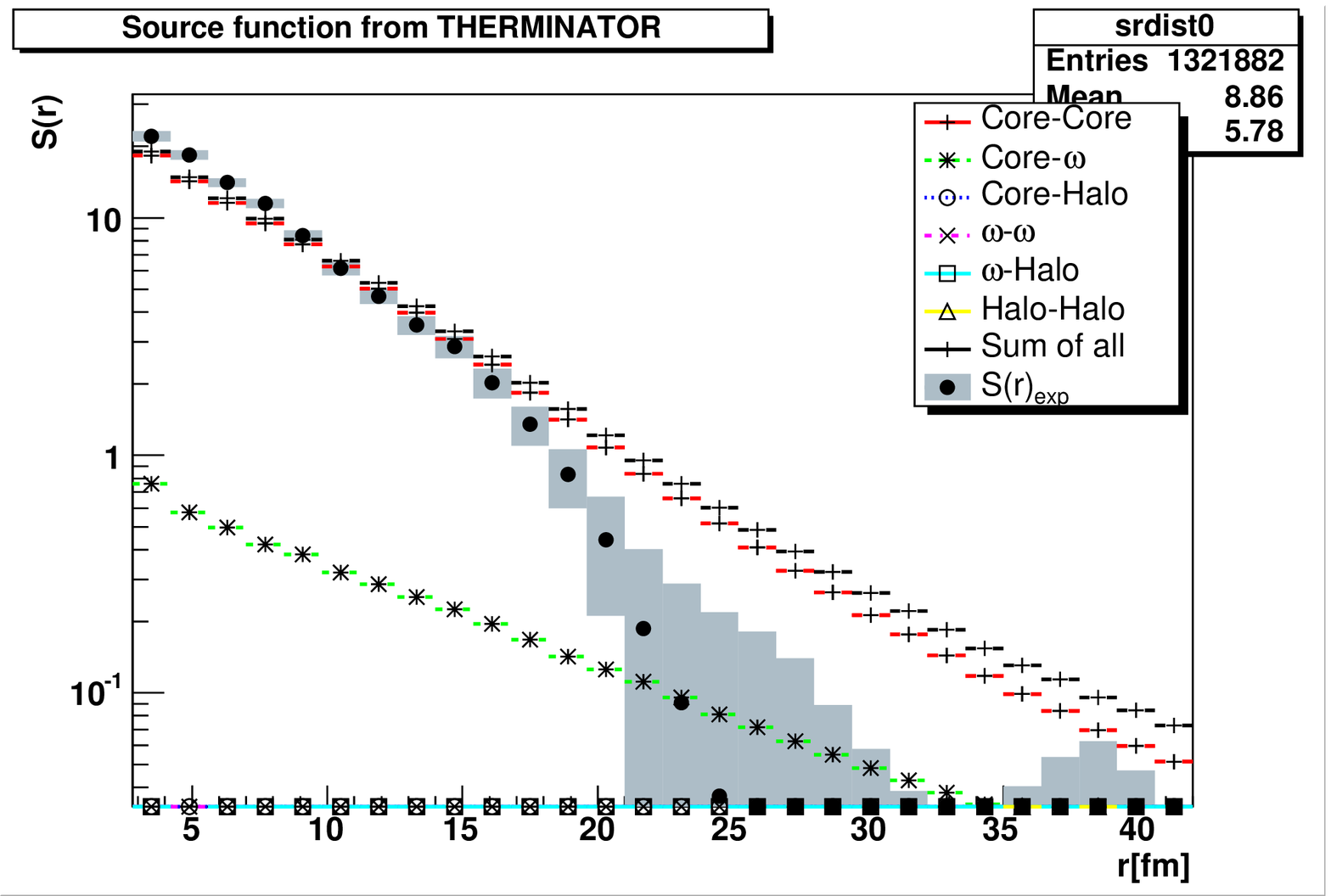} 
  \\
  \includegraphics[width=55mm]{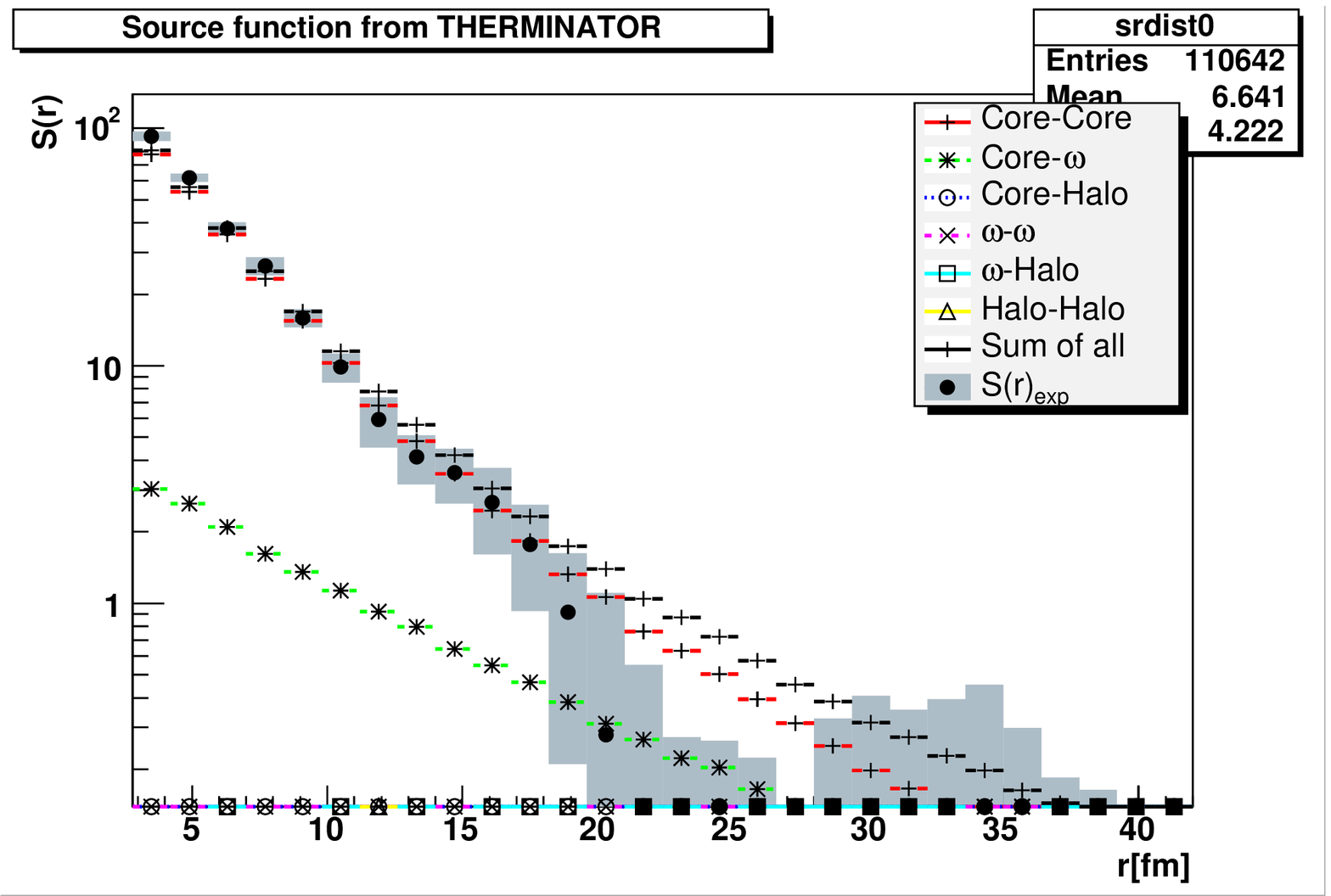}
  \caption[]{\it Pion pair source measured in PHENIX and THERMINATOR simulation for central
    ($0-20\%$) \kTlo collisions (top left); high \kT{} (\kThi) from the 20\% most
    central Au+Au midy collisions (top right); low kT (\kTlo) from
    peripheral collisions ($40-90\%$) (bottom center).
    Solid dots represent data on each figure. Black crosses are simulation, while
    color online symbols stand for different
    components of the simulated source as explained on the legend.}
\label{fig:th}
\end{center}
\end{figure}

A separate set of events were simulated for each centrality bin of
5\% for central, and 10\% for peripheral collisions. Then these
sets were concatenated to make up the wider regimes of PHENIX
data. THERMINATOR was able to reproduce the centrality dependence
of the data. It is also capable to describe the \kT{} dependence
to a much higher accuracy than HRC does, although there is a
significant discrepancy at the higher \kT{} region. (See Fig.\
\ref{fig:th}).

\begin{figure}[htb]
\begin{center}
  \includegraphics[width=55mm]{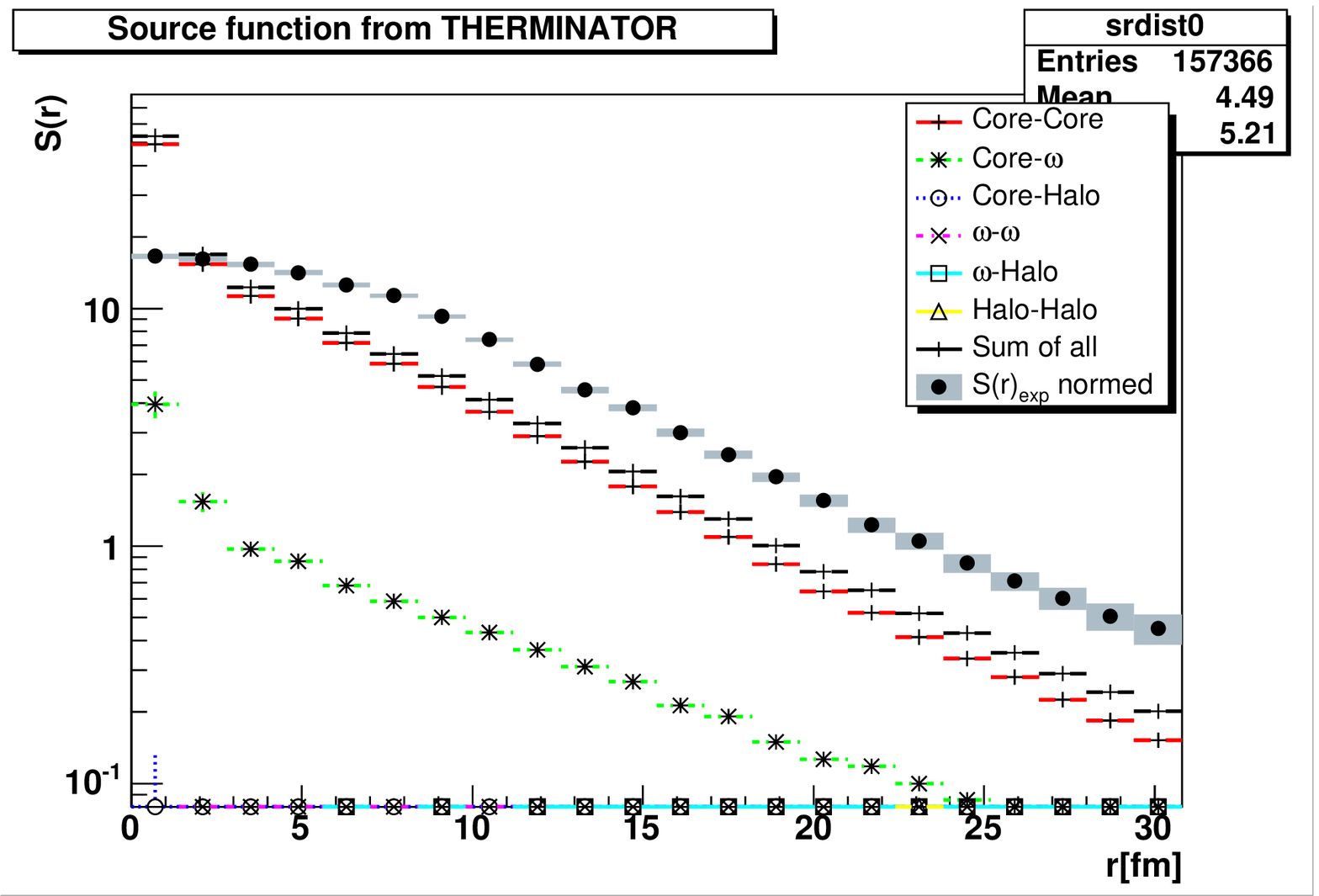}
  \includegraphics[width=55mm]{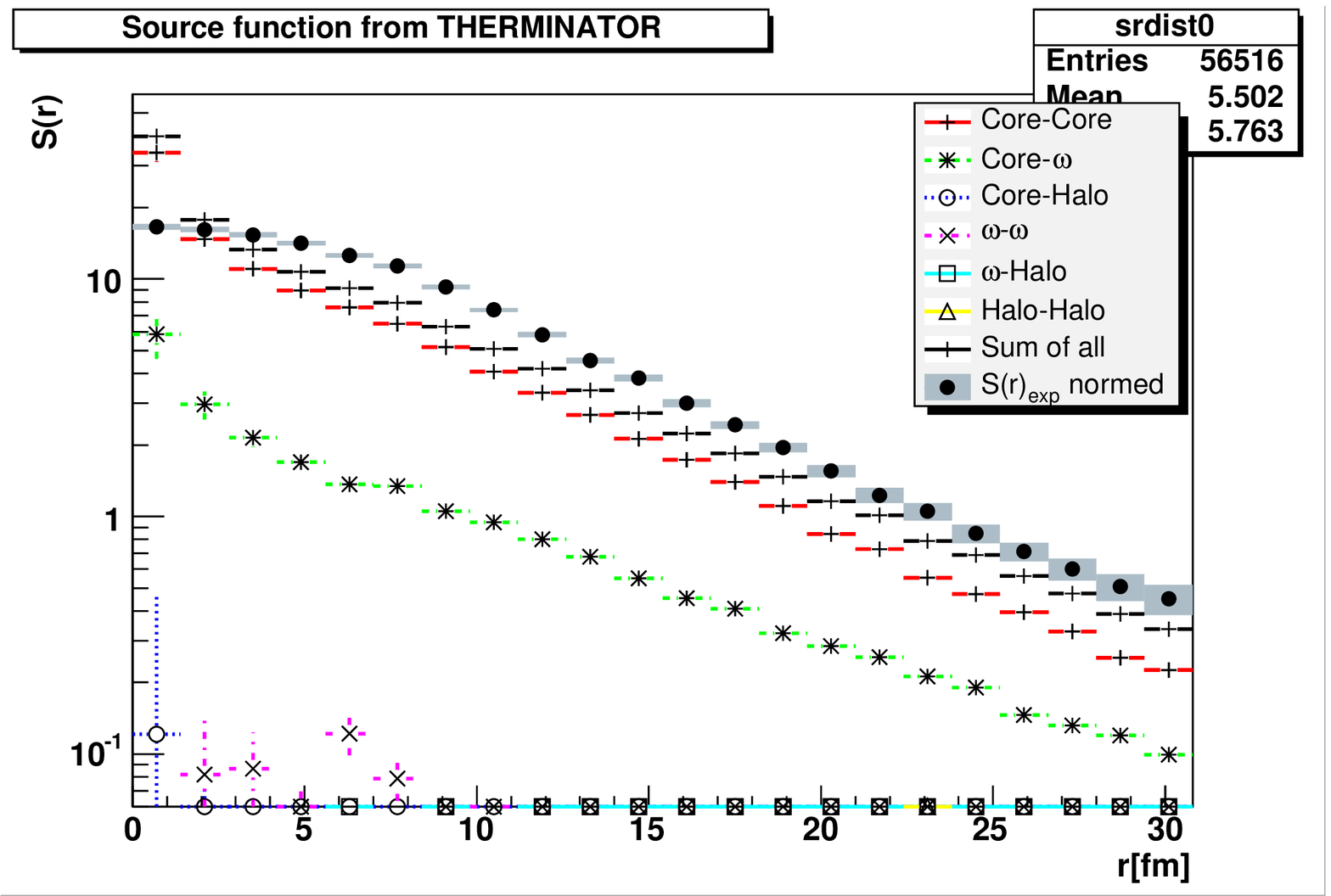}
  \\
  \includegraphics[width=55mm]{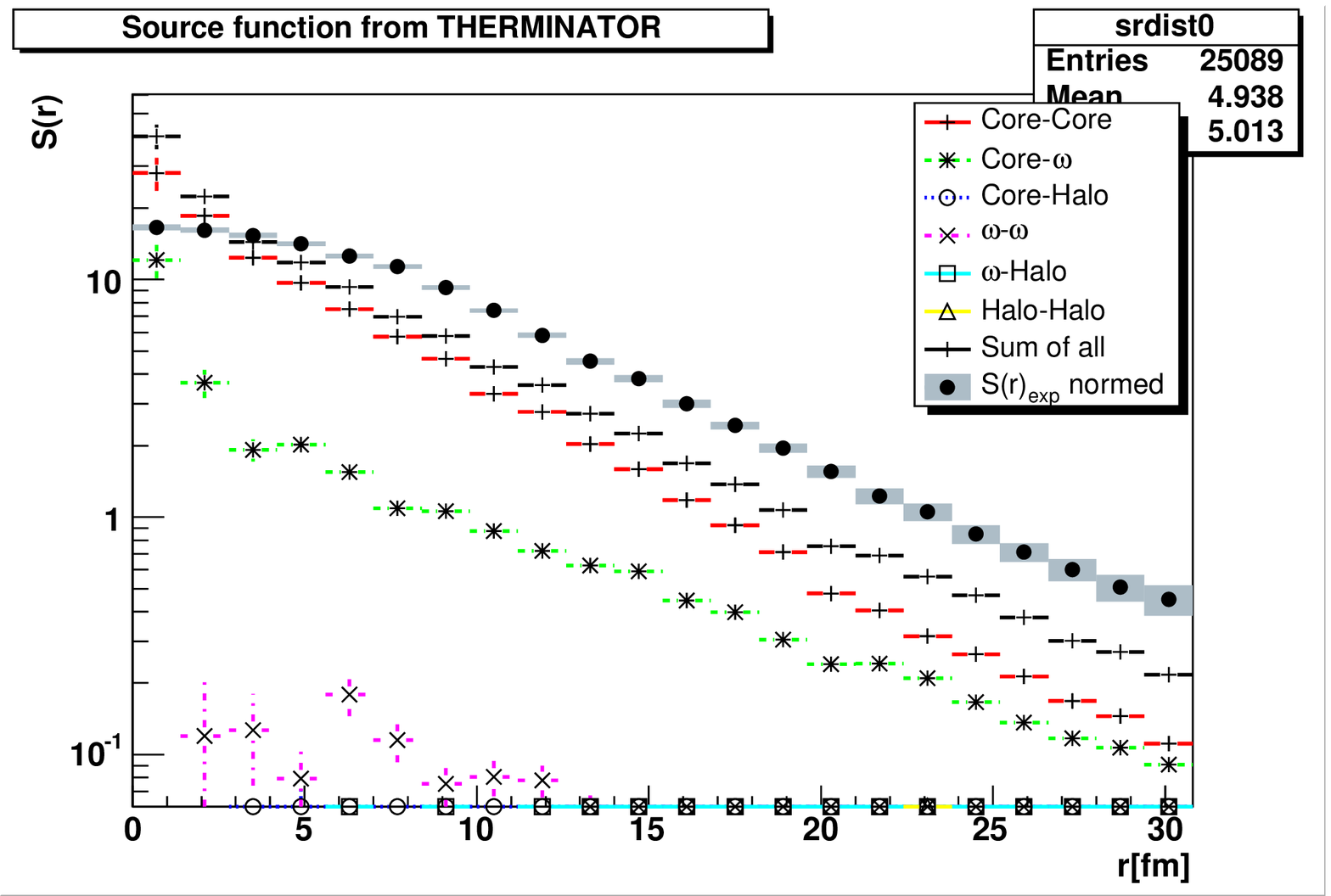}
  \includegraphics[width=55mm]{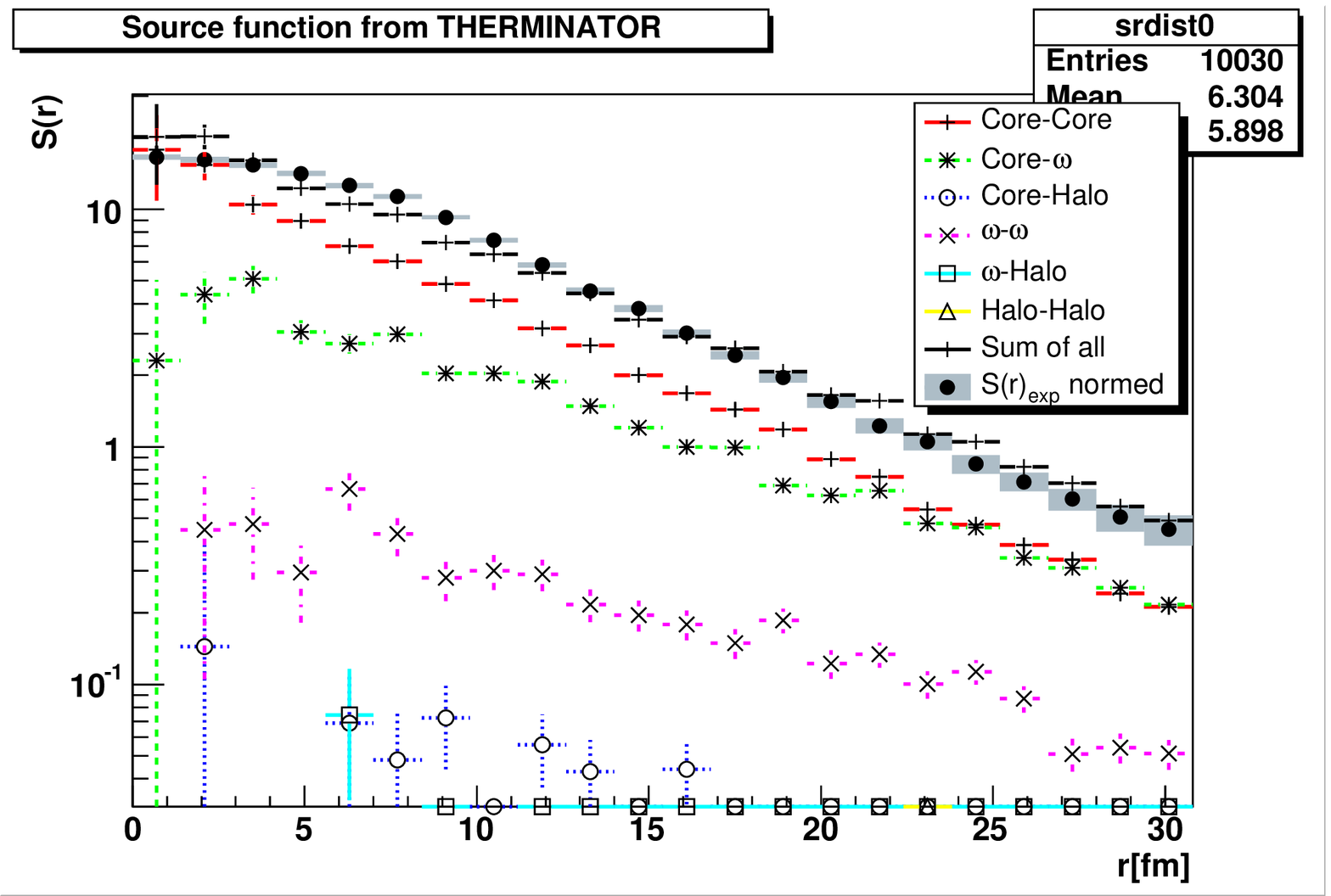}
  \caption[]{\it $r\rightarrow{}0$ behaviour of THERMINATOR simulated \pip\pip{} and \pim\pim{} \Sr{} for central
    low \kT collisions. Plots from top left to bottom right show all pion pairs, all pairs from resonance,
    all primordial pairs, and all pairs coming from the short lived resonances
    $\Gamma\geq{}150\MeV$ respectively. Solid dots represent data on each figure.
    Black crosses are simulation, while the color online symbols stand for different
    components of the simulated source as explained on the legend.}
\label{fig:th_r_studies}
\end{center}
\end{figure}

It must be noted that there is a significant excess in the
simulated \Sr{} at small $r$ values. Our studies have shown that
neither the primordial nor the resonance particles are responsible
for it alone, but the $r\rightarrow{}0$ peak vanishes if all the
core particles and the short lived ($\Gamma\geq{}150\MeV$)
resonances are excluded from the source (see Fig.\
\ref{fig:th_r_studies}). This can be explained with the freezeout
that occurs in a single surface in this model, and the lack of
rescattering.

Further simulations indicate that rescattering and resonance decay create
very different sources for different particles ($\pi^\pm, \mathrm{K}^\pm$ and 
$\mathrm{p}^\pm$).

\vspace{5mm}
\noindent
A study of direction-dependent PHENIX pion-pair source functions
using THERMINATOR Blast Wave and Single Freezeout and models has
been carried out by P. Chung et al.\ \cite{wwnd_pd} and concluded
that these models, although they do well in the "sideways" and
"long" directions, fail in the "outward" direction 
(in the terms of the Bertsch-Pratt parametrization \cite{Pratt}).

\subsection{Discussion of results}
THERMINATOR is a simple model that shows a high power in the
description of particle spectra of the large heavy ion
experiments. Although there is no rescattering implemented in this
model, it reproduces the PHENIX image with the observed heavy
power-law tail for the lower \kT~events with a higher accuracy
than HRC---although it still slightly overpredicts the tail for the high
\kT~regime, and fails for $r\rightarrow{}0$ because of its
inherent limitations.

The relatively good agreement including the reproduction of the long-range tail
can be understood by taking into account that the high number of
resonances provide us with an almost-continuum distribution of
lifetimes, increasing virtually to infinity. This results in an approximate
power-law-like final distribution \cite{Bialas:1992ca} just as anomalous diffusion
does \cite{Metzler_Klafter}. However, THERMINATOR fails to describe correctly the 
intercept $\lambda$ and the HBT radii $R_{side}, R_{long}$ and $R_{out}$ 
\cite{Warsaw_picorr}. 
So it is questionable that resonance decays alone were sufficient
to describe the experimentally seen structure of source as Bialas had proposed.

\section{Conclusions}\label{lbl:concl}
Source images of PHENIX data have been studied with two fundamentally different,
simple models. Both were able to reproduce the observed 1D images
to a certain level, but had their limits. The different kinematic
mechanisms give similar shapes, that are determined by the same
underlying mathematical principle (a generalization of the Central Limit Theorem),
and so that any of them can be a potential explanation of the measured heavy power-law 
tails.

\begin{enumerate}
\item HRC simulations describe low \kT~PHENIX image, but do
not have the right \kt~dependence, as for higher \kT~values, the
simulated \Sr12~still has the large tail, but not the PHENIX data.
The reason for the heavy tail is rescattering, change of mean free
path and thus anomalous diffusion.

\item THERMINATOR simulations provide a reasonable reproduction of the \kT~and centrality 
dependent behaviour of the 1D PHENIX image. The heavy tail in those simulations can be 
explained by the large number of resonances in that model. This produces 
an effect that is similar to the anomalous diffusion.

\item However, THERMINATOR fails to describe the direction-dependently
parametrized PHENIX source functions, suggesting that resonance decays
alone are not a sufficient explanation of the measured structure of the source.

\item Measurement of $S(r)$ images for $\pi^\pm, \mathrm{K}^\pm$ and 
$\mathrm{p}^\pm$ could distinguish between rescattering and resonance decay effects.

\end{enumerate}


\vfill\eject

\begin{thebibliography}{9}

\bibitem{HBT}
R.~Hanbury Brown and R.~Q.~Twiss,
Nature {\bf 178}, 1046 (1956).

\bibitem{Koonin}
S. E. Koonin, 
Phys. Lett. B70, 43 (1977)

\bibitem{Messiah}
A. Messiah, ``Quantum Mechanics. Vol. 2'',
De Gruyter, Berlin (1979).

\bibitem{Alt}
E.~O.~Alt, T.~Cs\"org\H{o} B.~L\"orstad and J.~Schmidt-S{\o}rensen,
arXiv:hep-ph/0103019.

\bibitem{PPG052}
  S. S. Adler et al. (PHENIX Collaboration),
  Phys.Rev.Lett.98:132301 (2007) 
  [arXiv:nucl-ex/0605032].
\bibitem{ImgSrc}
D. A. Brown et al., 
Phys. Lett. B398, 252 (1997).

\bibitem{Humanic:2002iw}
  T.~J.~Humanic,
  Nucl.\ Phys.\ A {\bf 715}, 641 (2003)
  [arXiv:nucl-th/0205053].

\bibitem{Humanic:2006ve}
  T.~J.~Humanic,
  Phys.\ Rev.\ C {\bf 73}, 054902 (2006)
  [arXiv:nucl-th/0602027].

\bibitem{Metzler_Klafter}
  R. Metzler and J. Klafter,
  Physics Reports {\bf 339}, 1-77.

\bibitem{0702032}
  M. Csan\'ad and T. Cs\"org\H{o}, 
  arXiv:hep-ph/0702032

\bibitem{Therm}
A. Kisiel, T. Taluc, W. Broniowski, W. Florkowski,
Comp. Phys. Comm. 174 (2006) 669 
[arXiv: nucl-th/0504047v2].

\bibitem{Cracow}
W. Broniowski and W. Florkowski,
Phys. Rev. Lett. 87 (2001) 272302 
[arXiv: nucl-th/0106050].

\bibitem{Cent}
A. Baran, W. Broniowski, W. Florkowski,
Acta Phys. Polon. B35 (2004) 779 
[arXiv: nucl-th/0305075].

\bibitem{wwnd_pd}
P. Danielewicz, "Analysis of Emission Shapes",
talk on the 23rd Winter Workshop on Nuclear Dynamics

\bibitem{Pratt}
S. Pratt,
Phys. Rev. D 33, 72 (1986).

\bibitem{Bialas:1992ca}
A.~Bialas,
Acta Phys.\ Polon.\ B {\bf 23}, 561 (1992).

\bibitem{Warsaw_picorr}
W. Florkowski, W. Broniowski, A. Kisiel, J. Pluta,
arXiv:nucl-th/0609054v1

\end{thebibliography}
\end{document}